\documentclass[conference]{IEEEtran}
\usepackage{cite}
\usepackage{amsmath,amssymb,amsfonts}
\usepackage{algorithmic}
\usepackage{booktabs}
\usepackage{graphicx}
\usepackage{textcomp}
\usepackage{xcolor}
\usepackage{caption}
\usepackage{listings}
\usepackage{subcaption}
\usepackage{wrapfig}
\usepackage[hidelinks]{hyperref}
\def\BibTeX{{\rm B\kern-.05em{\sc i\kern-.025em b}\kern-.08em
    T\kern-.1667em\lower.7ex\hbox{E}\kern-.125emX}}

\usepackage{color}
\definecolor{mygreen}{rgb}{0,0.6,0}
\definecolor{myblue}{RGB}{49, 130, 189}
\definecolor{mygray}{rgb}{0.5,0.5,0.5}
\definecolor{mymauve}{rgb}{0.58,0,0.82}
\newcommand{\mopt}{\includegraphics[height=0.7em,page=1]{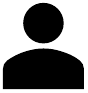}}
\newcommand{\aopt}{\includegraphics[height=0.7em,page=2]{figures/icons.pdf}}

\begin{document}

\title{Productive Performance Engineering for Weather and Climate Modeling with Python}

\author{
\IEEEauthorblockN{Tal Ben-Nun$^*$, Linus Groner$^\dagger$, Florian Deconinck$^\ddagger$, Tobias Wicky$^\ddagger$, Eddie Davis$^\ddagger$, Johann Dahm$^\ddagger$,\\Oliver D. Elbert$^\ddagger$, Rhea George$^\ddagger$, Jeremy McGibbon$^\ddagger$, Lukas Tr\"umper$^*$, Elynn Wu$^\ddagger$,\\Oliver Fuhrer$^\ddagger$, Thomas Schulthess$^\dagger$, Torsten Hoefler$^*$}
\IEEEauthorblockA{* \textit{Department of Computer Science, ETH Zurich}, Switzerland\\
$\dagger$ \textit{Swiss National Supercomputing Centre}, Switzerland\\
$\ddagger$ \textit{Allen Institute for Artificial Intelligence}, WA, USA\\
\{talbn, lukashans.truemper, htor\}@inf.ethz.ch, \{linus.groner, schulthess\}@cscs.ch,\\\{floriand, tobiasw, eddied, johannd, olivere, rheag, jeremym, elynnw, oliverf\}@allenai.org}
}

\maketitle

\global\csname @topnum\endcsname 0
\global\csname @botnum\endcsname 0

\begin{abstract}
Earth system models are developed with a tight coupling to target hardware, often containing specialized code predicated on processor characteristics. This coupling stems from using imperative languages that hard-code computation schedules and layout.
We present a detailed account of optimizing the Finite Volume Cubed-Sphere Dynamical Core (FV3), improving productivity and performance.
By using a declarative Python-embedded stencil domain-specific language and data-centric optimization, we abstract hardware-specific details and define a semi-automated workflow for analyzing and optimizing weather and climate applications.
The workflow utilizes both local and full-program optimization, as well as user-guided fine-tuning.
To prune the infeasible global optimization space, we automatically utilize repeating code motifs via a novel transfer tuning approach.
On the Piz Daint supercomputer, we scale to 2,400 GPUs, achieving speedups of up to 3.92$\times$ over the tuned production implementation at a fraction of the original code.
\end{abstract}

\section{Introduction}
\label{sec:intro}

\begin{figure}[t]
\centering
\includegraphics[height=1.65in,page=1]{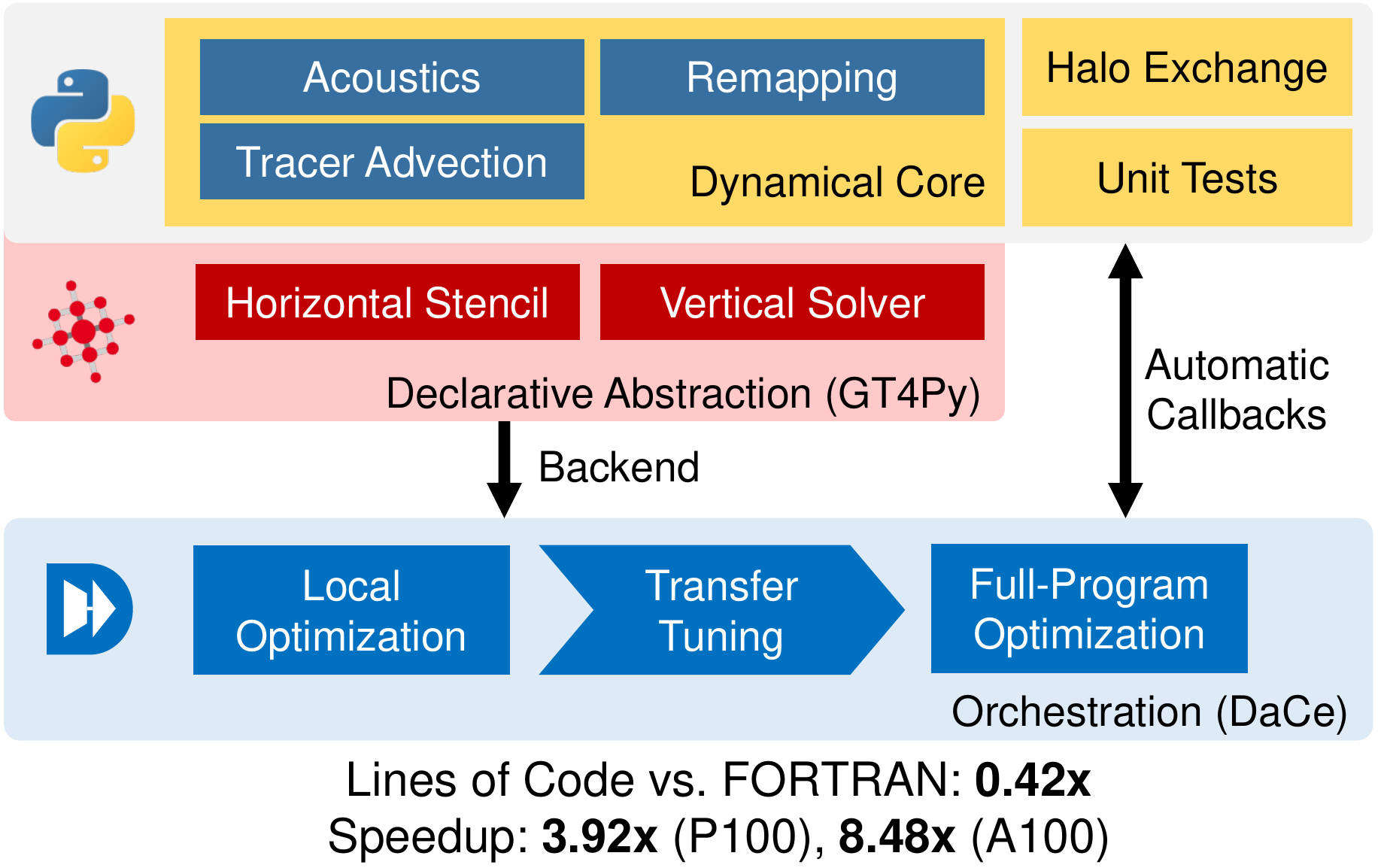}
\caption{System overview.}
\vspace{-1.5em}
\label{fig:system}
\end{figure}

Climate change and the associated weather extremes is one of the biggest challenges facing humanity today.
The basis of what we know about our future stem from simulations using weather and climate models running on some of the world's largest supercomputing infrastructures.
But progress in leveraging the power of current and emerging supercomputing hardware architectures is slow, due to legacy software engineering and lift-and-shift approaches to code portability.

In order to address the shortcomings of currently available weather and climate data, there is an urgent need to address this software productivity gap to enable simulations with higher fidelity~\cite{schulthess2018reflecting}.
Given today's hardware landscape and increasing complexity of models, the current approach is not sustainable~\cite{gmd-11-1799-2018}.
We urgently need to enable more rapid development cycles and faster adoption of leadership class supercomputing infrastructures~\cite{bauer_nature}.

In this paper, we present a novel approach to productive performance engineering for weather and climate modeling using Python. Our approach, summarized in Fig.~\ref{fig:system}, increases developer productivity while not making any compromises in terms of performance and performance portability.

To enable performance engineering, we must express the code in a way that allows it to mutate schedules (i.e., work distribution among processors and order of operations) while maintaining the algorithms. To this end, we leverage a domain-specific language (DSL) embedded in Python, called GridTools for Python (GT4Py), which allows the developer to express the algorithms on a high level of abstraction. The Python programming language and package ecosystem allows for writing modular, re-usable code which can easily be unit-tested (Section~\ref{sec:implementation}). The pure-Python backend of the DSL toolchain is ideal for rapid prototyping, debugging and interactive visualization of algorithmic approaches.
The abstraction and domain-specificity of the DSL allows for concise, declarative code which is not littered with hardware dependent optimizations or annotations.

For the actual performance engineering we leverage a data-centric (DaCe) parallel programming framework as a backend for optimization and code generation (Section~\ref{sec:dace} and~\ref{sec:mapping}). This framework allows to modify schedules, data layouts, and memory placement (Section~\ref{sec:optimization}). We take a disciplined model-driven approach for optimization~\cite{hoefler-sop-modeling}, which keeps track of performance bounds (e.g., memory bandwidth) during local and global optimizations to guide further decisions.

Furthermore, we introduce a novel automatic tuning technique called transfer tuning (Section~\ref{sec:optimization:transfer}).
The core problem of automatic tuning at scale is the high rate of configurations that must be traversed.
Since many optimizations recur throughout the program, good configurations are already known from earlier decisions.
Transfer tuning therefore extracts the top configurations for each type of transformation from a smaller part of the application and applies matching patterns to the others, significantly pruning the search space.

To illustrate our approach we choose the Finite-Volume Cubed-Sphere Dynamical Core (FV3), an open-source solver implemented in FORTRAN. FV3 is used as the workhorse of several popular community weather and climate models. It was specifically designed to reap the benefits of multicore CPUs~\cite{lin_vertically_2004, putman_finite-volume_2007} by keeping two-dimensional fields for most of the time step, thereby exhibiting high cache utilization.

We port FV3 to GPUs, whose microarchitecture and programming model do not share the same cache benefits. With the DSL and global optimization, we gain a 3.9$\times$ speedup at scale on the Piz Daint supercomputer.
As the methodology applies to any model, it serves as a testament that Python-based models can be feasible, productive, and fast.

The paper makes the following contributions:
\begin{itemize}
    \item Methodology for guided weather and climate performance engineering, defining the domain-specific search space for local and global optimization.
    \item Performance modeling tools for said methodology to bound performance (i.e., indicate when to stop optimizing certain components) and guide tuning when using data-centric Python and declarative stencil definitions.
    \item \textit{Transfer Tuning}, a novel technique that drastically reduces auto-tuning cost in large applications by reusing beneficial patterns.
    \item Demonstrating the methodology on the FV3 weather model, where for the first time a high-performance, \textbf{full dynamical core} is entirely written in, compiled from, and run on up to 2,400 GPUs within Python. This includes a comprehensive account of the involved optimizations, performance upper bounds for the sizes used in operational forecasting, and effort estimation for porting the approach to new hardware or weather models.
\end{itemize}

\section{The Finite-Volume Cubed-Sphere (FV3) model}
\label{sec:fv3}

The \textit{dynamical core} is the foundation of any weather and climate model. It integrates the governing equations of air flow and thermodynamics forward in time from a given initial state. Finite-Volume Cubed-Sphere Dynamical Core (FV3) is a dynamical core developed at the Geophysical Fluid Dynamics Laboratory (GFDL) and has been chosen as the dynamical core for the Next Generation Global Prediction System project~\cite{ji_dynamical_2016}. FV3 has been deployed in numerous community models (e.g., SHiELD, GEOS, UFS, CESM) both in production and research and has shown promising performance in speed and accuracy on the global and regional scale~\cite{harris_gfdl_2020}.

While FV3 is capable of solving both the hydrostatic and non-hydrostatic set of governing equations, the focus of this paper will be on the non-hydrostatic option. In particular, FV3 solves the fully compressible Euler equations on the gnomonic cubed-sphere grid and a Lagrangian vertical coordinate~\cite{harris_two-way_2013, lin_vertically_2004, putman_finite-volume_2007}. It is fully explicit in the time integration except for fast vertically propagating sound and gravity waves. FV3 is designed to be computationally efficient and many algorithms have been selected or revised for optimal efficiency. In fact, efficient implementation of the scientific algorithms is one of the defining traits of FV3. The reference of FV3 is implemented in FORTRAN and parallelized using OpenMP directives for on-node threading and a two-dimensional domain decomposition in the horizontal dimensions using MPI library calls.

\begin{figure*}[t]
    \centering
    \includegraphics[width=\linewidth]{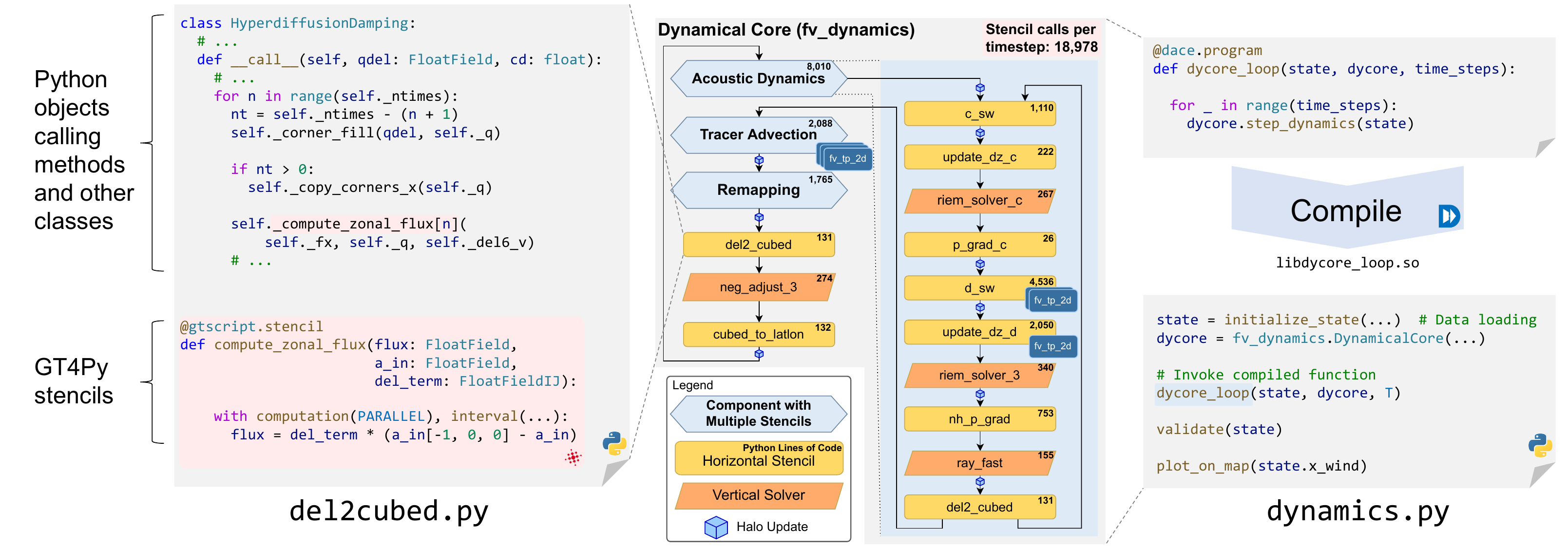}
    \caption{Code structure of the FV3 dynamical core. The entire codebase is written in object-oriented Python, where stencils use the GridTools for Python (GT4Py) embedded DSL. The time-stepping loop is then compiled to a library using the DaCe framework, and invoked as a single call to avoid Python interpreter overhead.}
    \vspace{-1em}
    \label{fig:fv3}
\end{figure*}

FV3 utilizes sub-stepping to integrate different processes on different timesteps. There are three levels of sub-stepping:
the \textbf{physics timestep} is the outermost loop and advances the model by an atmospheric timestep in every iteration.
Each iteration contains calls to the dynamical core as well as potential calls to desired physical parametrizations.
As this work is focused on the dynamical core, we leave any physics packages out of the analysis.
The \textbf{remapping timestep} is the middle loop found in the model (see Fig.~\ref{fig:fv3}, middle, for an overview of the structure of the remapping timestep).
The number of substeps can be tuned to balance performance and model stability.
In each remapping timestep the model advects tracer variables (middle hexagon in Fig.~\ref{fig:fv3}), remaps the deformed Lagrangian to the reference ``Eulerian'' vertical coordinate levels (bottom hexagon) and contains a call to the innermost loop, where the integration of the dynamical equations happens on Lagrangian surfaces.
The number of iterations of the innermost loop of the model, the \textbf{acoustic substep} (top hexagon), is determined by horizontal sound wave propagation and thus directly related to the grid spacing. In this innermost loop, there are several execution points where nonblocking halo exchanges of one or more fields occur.

The FORTRAN implementation of the model has been heavily optimized for multi-core CPU architectures. The structure of the solver lends itself to leveraging data-locality in the cache hierarchy of the CPU, since large blocks of code act solely on horizontal planes of the data fields. In the strong scaling regime, sub-domains assigned to a single MPI process are small enough so multiple two-dimensional horizontal planes fit into an L2 cache. In these parts of the code (many of the yellow boxes in Fig.~\ref{fig:fv3}), the vertical loop has been hoisted outwards as far as possible (in a strategy called \texttt{K}-blocking). This strategy is a good choice for this specific dynamical core as there is very limited vertical coupling throughout the model.

As part of the first intercomparison project of global storm-resolving models (GSRMs) – the DYnamics of the Atmospheric general circulation Modeled On Non-hydrostatic Domains (DYAMOND) \cite{stevens_dyamond_2019} – FV3 demonstrated the capability of running a global simulation at 3.3 km horizontal resolution at 19 simulation days per day using 384 nodes (dual-socket, Intel Broadwell, 18 cores/socket) and produced remarkable similarities for a snapshot comparison against satellite observations from Himawari 8, making itself a top contender of GSRMs in the world.

\section{FV3 Optimization Efforts}
\label{sec:py-fv3}

Recently, the push towards a km-scale climate model is advocated by the community to commensurate with the challenges posed by climate change \cite{palmer_scientific_2019}. High resolution simulation has the ability to better represent weather phenomena such as deep convection, which remains a source of uncertainty in climate predictions.
However, it is difficult to run such simulations at a high enough throughput without major change in programming paradigms (i.e., the ability to run climate models efficiently on current and emerging computer hardware with hybrid nodes)~\cite{schulthess2018reflecting}.
To this end, several directions were researched in making FV3 performance portable. 

\subsection{Domain-Specific Languages for Weather and Climate}

While parts of FV3 were optimized for other hardware targets, the full model was not ported to a different architecture. Several efforts~\cite{fv3-gpu,parallel-fv3} have shown that parts of FV3 can run faster on GPUs than CPUs using CUDA FORTRAN~\cite{cudafortran} or compiler directive (e.g., OpenACC) approaches. This, however, creates hardware-specific code that is unlikely to run with good performance on other architectures, and is difficult to read and maintain when compiler libraries and requirements change. Additionally, the prevalent algorithmic motifs in atmospheric models are a limited subset of numerical operations, mostly consisting of stencil computations on 3D arrays and columns, and thus are well suited to specializing optimizations that can outperform general-purpose compilers.

Using a DSL expands the possibilities for both how to express and how to optimize FV3 code. The three P’s (productivity, portability, performance) are aspirational concepts for application sustainability \cite{three-ps} that DSLs can readily address. By creating a layer of abstraction to represent the algorithms of the model separate from the code optimized for a specific hardware, the abstracted code can be expressed in a way that is similar to the discretized mathematical equations, so that working on the algorithms is straightforward.

Several DSLs targeting atmospheric applications demonstrate the validity of this approach. CLAW \cite{claw} is a FORTRAN-based DSL supporting patterns commonly found in weather and climate applications and has been shown to be effective for certain parts of weather and climate models.
DAWN \cite{Dawn_2020} and the Open Earth Compiler \cite{open-earth} are compiler toolchains for weather and climate applications that provide abstractions between a high level expression of algorithm while generating efficient code for multiple platforms.
Julia has also been used to construct models~\cite{schneider2017earth}.
As for feasibility of using a DSL for an entire model, the UK Met Office recently demonstrated deploying an entire dynamical core over the LFRic toolchain~\cite{lfric}.

\subsection{Why Python as the driver language}

Python is a popular choice of language in Earth system sciences, particularly for processing, analyzing and visualizing Earth system model output.
Python has many well-documented libraries available for this type of analysis, ranging from machine learning libraries such as Scikit-learn \cite{scikit-learn} and TensorFlow \cite{abadi2016tensorflow} to domain-specific libraries such as xarray \cite{hoyer2017xarray} or Cartopy \cite{Cartopy}.
Additionally, its popularity in introductory programming courses make it have a very low barrier of entry for most new researchers in their standard workflow.

However, Earth system models are generally not written in Python.
Data are typically saved to the filesystem by a compiled language such as FORTRAN before being post-processed in a Python script.
Writing data to and reading it from the filesystem can be limiting due to storage constraints, and increases processing time.
Furthermore, modifying the behavior of the model often means writing code in the compiled language, restricting the ability to use Python libraries.
This is particularly an issue when trying to improve Earth system models with online machine learning, given the popularity of Python for training learned models \cite{mcgibbon2021fv3gfs}.

Earth system scientists have previously used multiple approaches to circumvent interoperability issues, including FORTRAN libraries to pass data between FORTRAN and Python \cite{SmartSim, PyFort}, wrapping a FORTRAN model as a Python library~\cite{monteiro2018sympl, mcgibbon2021fv3gfs}, writing FORTRAN libraries that can run Python-trained machine learning models \cite{ParallelFortran,ott2020fortran}, or writing the model itself in Python or Cython\cite{pressel2015large,lin2009qtcm, ClimaLab}.

\subsection{Python-driven FV3}
To circumvent issues with Python/FORTRAN interoperability, we chose to rewrite the FV3 dynamical core and its driver code entirely in Python, as summarized in Fig.~\ref{fig:fv3}.
Each of the FV3 modules is situated in a class, which invokes stencils via a stencil DSL, as well as other modules that may contain stencils of their own (Fig.~\ref{fig:fv3}, left).
The driver code is then automatically generated as C++/CUDA code and compiled to a shared library. The library handles all aspects, from memory management to distributed communication, and seamlessly integrates with the Python interpreter without changing the function signature (Fig.~\ref{fig:fv3}, right).
This approach allows us to combine the flexibility and productivity of a scripted language with the performance gains from a compiled language.

\section{Workflow Components}
\label{sec:infrastructure}

To support our pure Python-driven approach, we use and extend the two Python-based frameworks described below.

\begin{figure}[t]
\centering
\begin{subfigure}[b]{\linewidth}
 \centering
 \includegraphics[width=\textwidth]{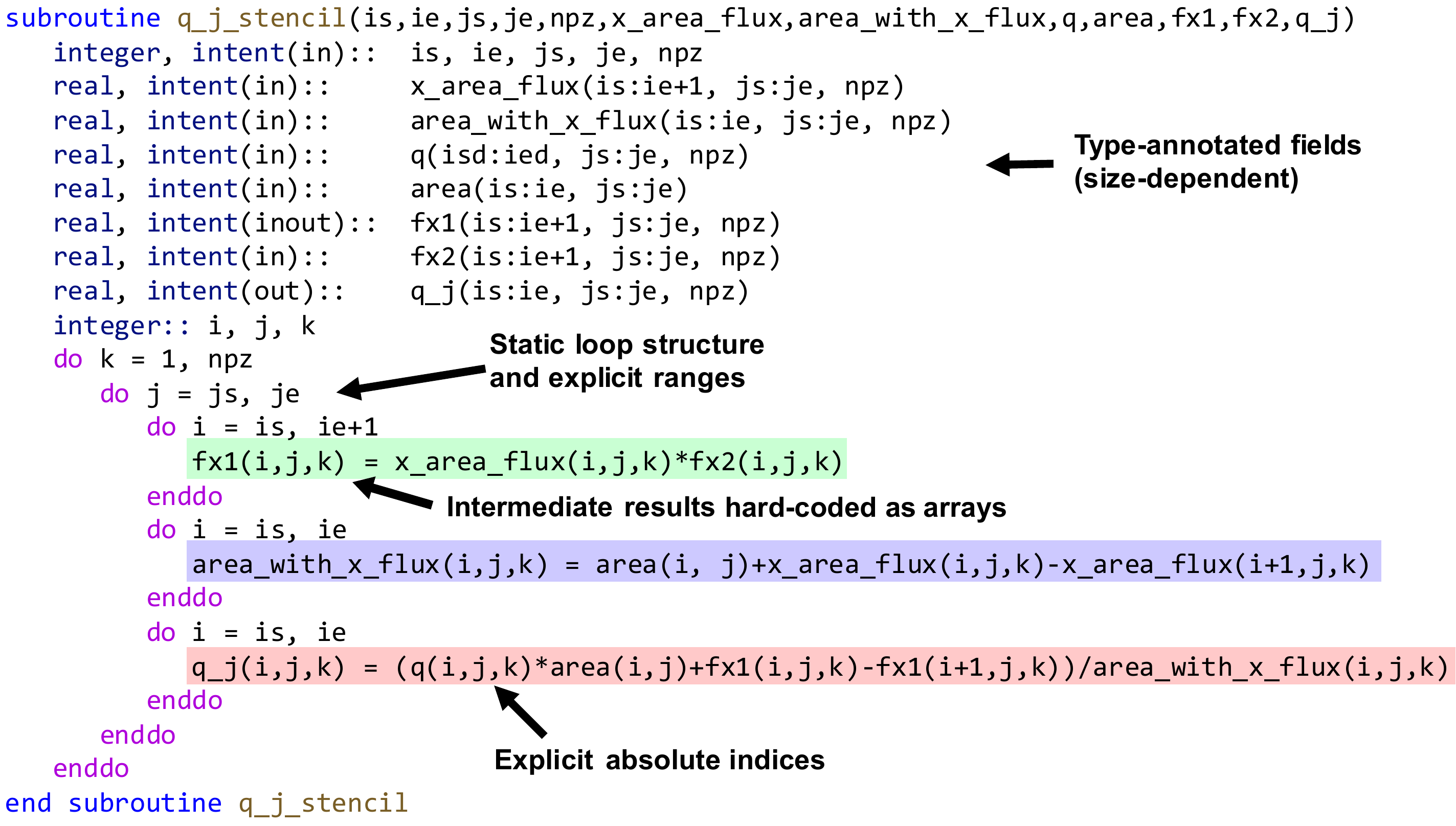}
 \caption{FORTRAN}
 \label{fig:code:fortran}
\end{subfigure}\vspace{1em}
\begin{subfigure}[b]{\linewidth}
 \centering
 \includegraphics[width=\textwidth,page=3]{figures/figures.pdf}
 \caption{GT4Py}
 \label{fig:code:gt4py}
\end{subfigure}
\caption{Stencil written in FORTRAN and its corresponding parallel stencil in GridTools for Python.}
\label{fig:code}
\vspace{-1em}
\end{figure}

\subsection{GridTools for Python (GT4Py)}
\label{sec:gt4py}

GT4Py~\cite{gt4py} is an embedded DSL, which enables fast authoring of stencil codes for Earth system models. In GT4Py, stencils are written in a declarative manner, abstracting away most of the hardware-specific constructs while keeping the code close to its original mathematical formulation. 

GT4Py stencils are written as Python functions with a {\small\texttt{@gtscript.stencil}} decorator (Fig.~\ref{fig:code}). In a decorated function, statements can be grouped with a {\small\texttt{computation}} block that defines the stencil type, while {\small\texttt{interval}} statement restricts the computations to the specified vertical intervals. Within a single block, each assignment corresponds to a single stencil operation. Authoring a GT4Py stencil is thus a relatively direct translation of the original code. More detailed semantics can be found in GridTools~\cite{gridtools}.

\begin{figure}[t]
\centering
\begin{subfigure}[b]{.45\linewidth}
 \centering
 \includegraphics[height=1in,page=4]{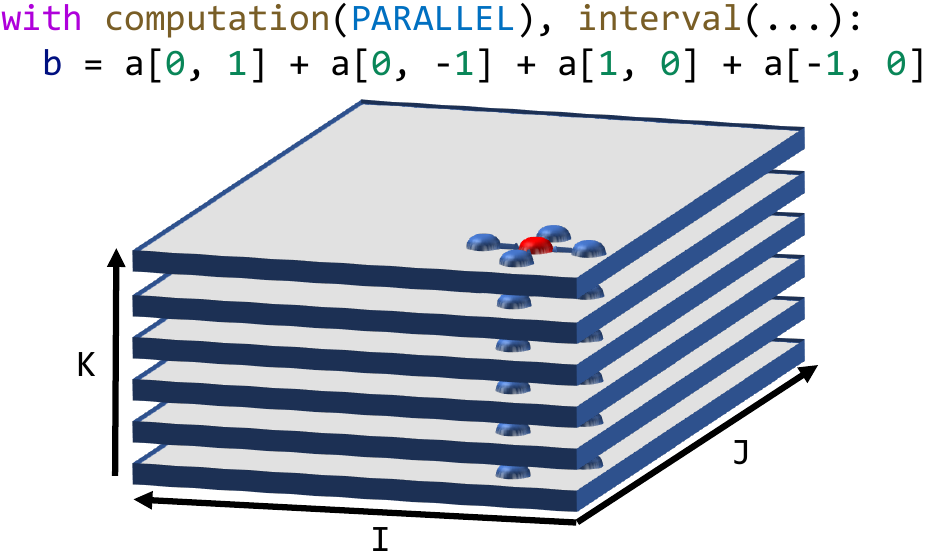}
 \caption{Horizontal Stencil}
 \label{fig:stencils:parallel}
\end{subfigure}\qquad
\begin{subfigure}[b]{.45\linewidth}
 \centering
 \includegraphics[height=1in,page=5]{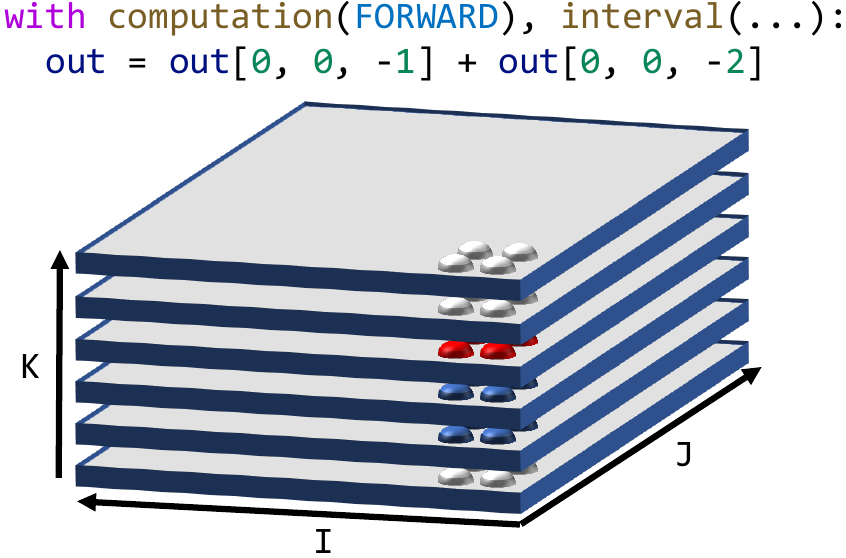}
 \caption{Vertical Solver (forward)}
 \label{fig:stencils:vertical}
\end{subfigure}
\caption{Horizontal and vertical stencil abstraction.}
\vspace{-1em}
\label{fig:stencils}
\end{figure}

GT4Py classifies stencils into two types, as depicted in Fig.~\ref{fig:stencils}. Horizontal (parallel) Stencils contain no loop-carried dependencies --- namely, the result of applying the stencil on one point cannot affect others --- and Vertical Solvers, in which the output field is computed in a vertical order and can use previously computed values in higher (\textit{forward} solvers) or lower (\textit{backward}) vertical levels.

GT4Py does not define a specific domain size for its stencils, only dimensionality (e.g., {\small\texttt{FloatFieldIJ}} for 2D horizontal fields), and all accesses to memory in stencil operations use relative indices. Buffer sizes for fields are thus transparently defined by inferring halo regions and extents from usage in stencils. Loop order, data layout, and even fusion of multiple operations into one stencil are similarly abstracted away.
The specified \textit{backend} is responsible for optimization, memory allocation and layout, and code generation, enabling stencil-level decisions on scheduling.

\subsection{Data-Centric Parallel Programming (DaCe)}
\label{sec:dace}

The DaCe framework is an optimization infrastructure written in Python, which enables interactive analysis and optimization of data movement in programs. At the core of DaCe lies an intermediate representation (IR) called Stateful Dataflow Multigraphs (SDFG)~\cite{dace}.

\begin{figure}[t]
\centering
\begin{subfigure}[b]{\linewidth}
 \centering
 \includegraphics[width=\textwidth]{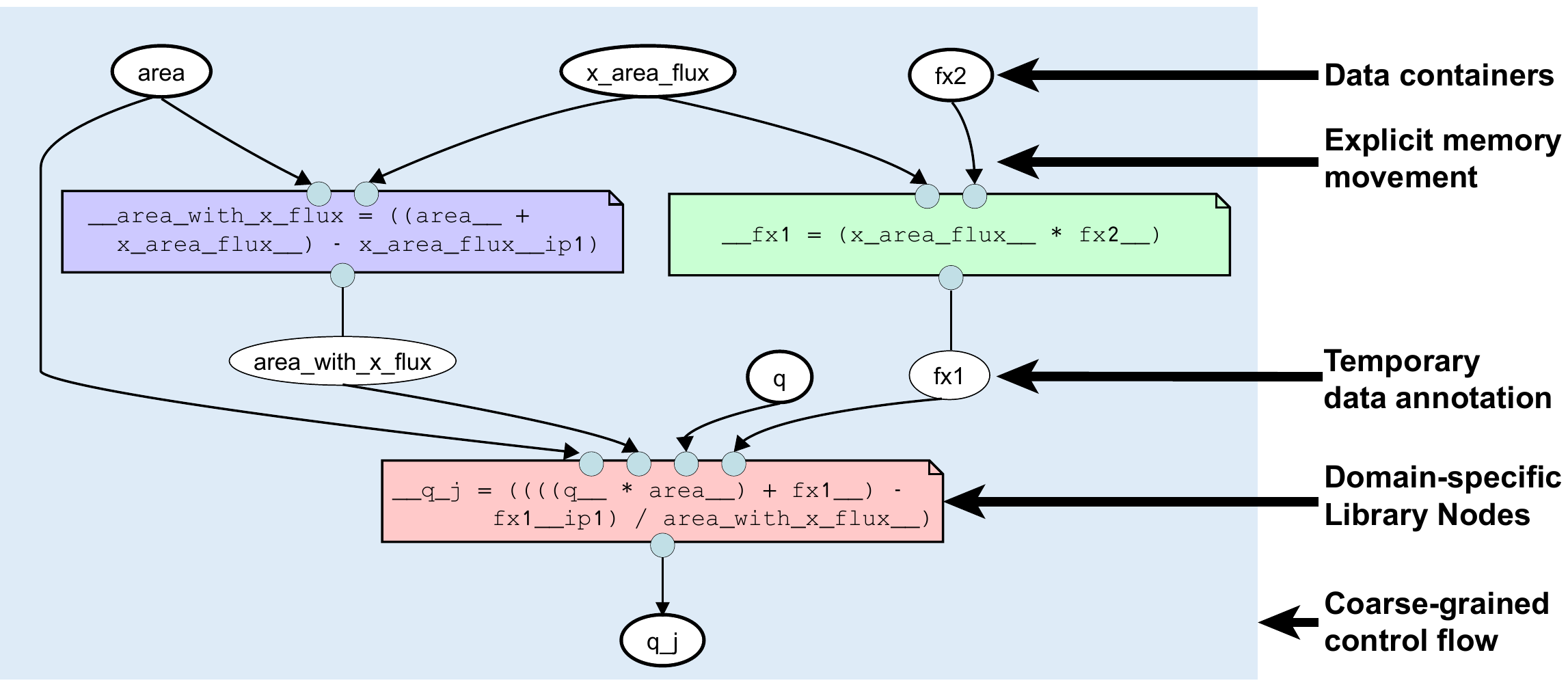}
 \caption{High-Level SDFG}
 \label{fig:code:sdfg-unopt}
\end{subfigure}\vspace{1em}
\begin{subfigure}[b]{\linewidth}
 \centering
 \includegraphics[width=.9\textwidth]{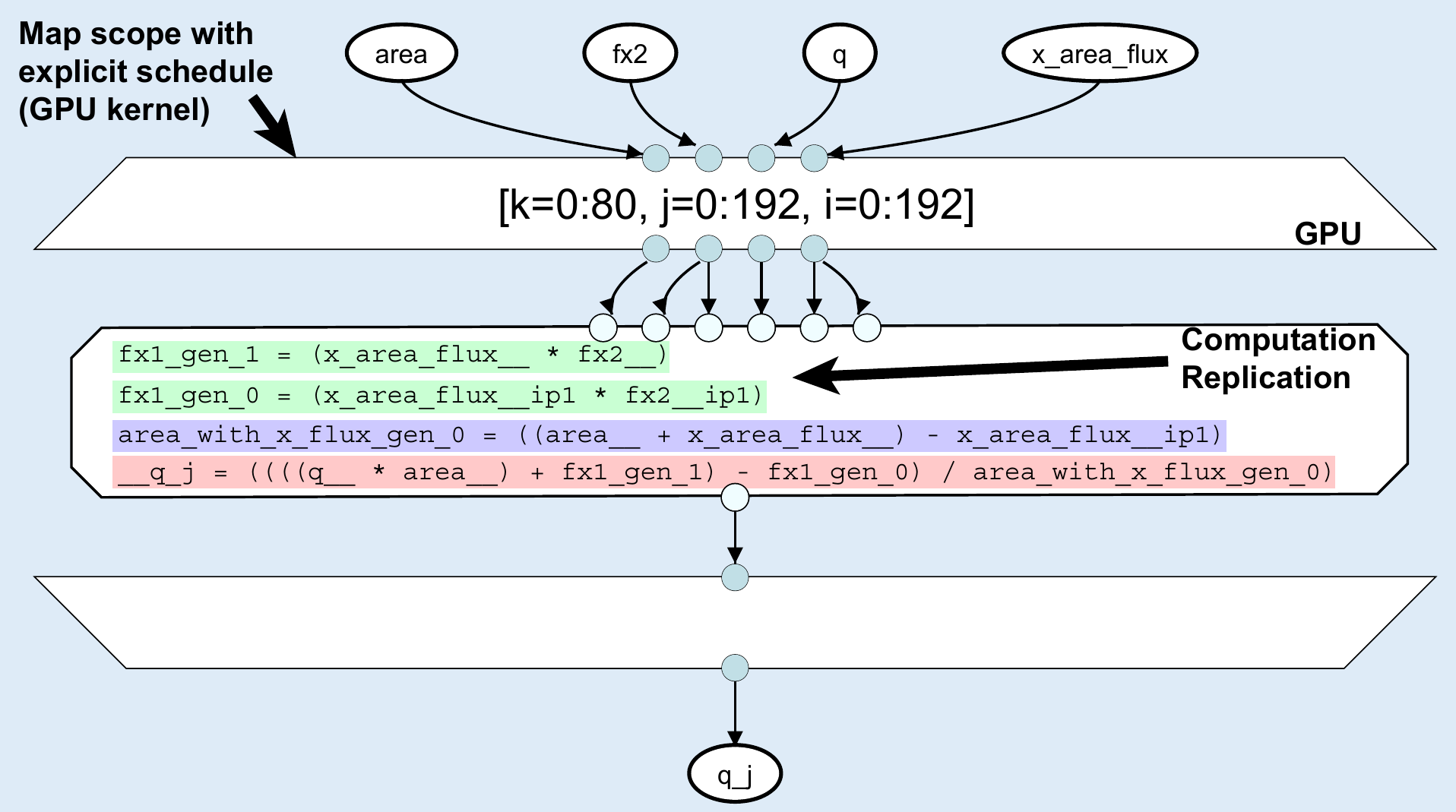}
 \caption{Expanded and Fused SDFG}
 \label{fig:code:sdfg-expanded}
\end{subfigure}
\caption{Stateful Dataflow Multigraph (SDFG) that corresponds to the stencil from Fig.~\ref{fig:code}.}
\vspace{-1em}
\label{fig:sdfg}
\end{figure}

The SDFG IR (Figures~\ref{fig:code:sdfg-unopt} and~\ref{fig:code:sdfg-expanded}) is composed of acyclic dataflow graphs nested in state machines, representing control flow. In each state, data containers (oval nodes) and data movement (edges) are explicitly defined separately from computations, which are represented by octagonal nodes. \textit{Map} scopes (trapezoidal sections) represent parametric parallelism, replicating the graph within them for the indicated domain. Coarse-grained domain-specific computations (e.g., matrix multiplication) can be represented by \textit{library nodes}, which can be expanded to ``native'' subgraphs containing the aforementioned components. This approach is widely used in DSLs that are lowered to SDFGs.

SDFGs inherently allow users to query data movement for exact ranges at any point of the program. This, in turn, enables \textit{data-centric optimizations} in the form of graph rewriting rules. Examples include removing redundant memory allocation, creating local storage, scheduling map scopes to run on accelerators, change code schedule arbitrarily (e.g., tiling, fusion), and others. Information on removable (\textit{transient}) containers is indicated on the graph.

DaCe can optimize programs written in different languages, including Python/NumPy~\cite{p3dace}, C~\cite{c2dace}, and embedded DSLs such as PyTorch~\cite{daceml}. The resulting SDFGs can then be optimized programmatically or interactively, and map to state-of-the-art code for a wide variety of hardware architectures, including multi-core CPUs, NVIDIA and AMD GPUs, and Xilinx and Intel FPGAs.

\section{Python FV3 Implementation}
\label{sec:implementation}

The dynamical core of FV3 suits the two stencil classes that GT4Py targets well, but not exactly. Thus, some design decisions and concessions had to be made on both the model and the DSL to allow for this approach to work.

\subsection{Object-oriented design}

To maintain productivity, the following features should be possible to use without any performance penalty: (1) Python classes with complex inheritance; (2) function calls at arbitrary depths; (3) Pythonic objects such as lists and dictionaries; and (4) use of arbitrary amounts of temporary variables without worrying about memory allocation. 
We chose to value both maintainability as well as productivity via object-oriented programming (OOP) that represents each module of the dynamical core as its own class. Classes have fields, methods with clear inputs and outputs, and the required subset of the global configuration of the model.

As for the class structure, we keep the structure of the original FORTRAN subroutines (see Fig.~\ref{fig:fv3}) as modules in the Python implementation. As the FORTRAN model was designed with an emphasis on separating different numerical steps of the model, there was no need to change that.

One significant new opportunity this approach enables is modularized, fine-grained testing.
We started the porting effort by saving input and output data for each of the target modules from the FORTRAN model.
Our test suite has independent standalone unit-tests for model validation by comparing with the stored reference up to a given numerical precision.
For stencils that do not have dependencies on other tiles of the cubed sphere, testing can even be performed on a single node in a ``sequential'' mode. All stencils have the option to be tested with any configuration of multiple subdomains on all six tiles supported by our standard partitioner.
This fine-grained testing allows model developers to get rapid feedback on model behavior with every change.

Apart from modularity, we can measure code complexity using lines of code as a proxy. A high-level comparison of model code sizes can be found in Table~\ref{tab:loc}. Even though this is not an apples-to-apples comparison, by virtue of the DSL toolchain, the Python version runs on a whole range of different hardware architectures, while the FORTRAN code only runs on multicore CPUs. While there are some differences between the codes (e.g., the FORTRAN version also supports hydrostatic modeling or nesting), there is an overall clear advantage to using Python, at 0.42$\times$ the code length.

\begin{table}[t]
    \centering
    \caption{Lines of Code (LoC) Comparison of FV3} \small
    \begin{tabular}{lcc}
    \toprule
        Module Name      & Python LoC & FORTRAN LoC \\\midrule
        Dynamical Core          & 12,450                   & 29,458                    \\
        Finite Volume Transport           & 686                     & 858                      \\
        Riemann Solver C & 253                     & 267                      \\
        \bottomrule
    \end{tabular}
    
    \label{tab:loc}
    \vspace{-1em}
\end{table}

\subsection{Horizontal Regions}
\label{sec:regions}

The cubed sphere grid on which FV3 runs contains correction terms that need to be applied at the corners and edges of each of the six tile faces in the grid.
We thus extend GT4Py to support specializing stencils to sub-regions in the horizontal domain, such as corners and edges.
For example, to specialize a computation on the lower edge in the second horizontal dimension, the syntax is as follows:

    {
        \footnotesize
        \begin{lstlisting}
with computation(PARALLEL), interval(...):
  flux = dt2 * (velocity - velocity_c * cosa) / sina
  with horizontal(region[:, j_start]):
    flux = dt2 * velocity
\end{lstlisting}
    }

If a stencil is distributed across multiple ranks in the horizontal plane, regions are also used to infer communication-related aspects. As only some of the regions read computed data from other ranks, the new DSL extension resolves which other ranks to synchronize with based on the ranges.

\subsection{Halo exchange}

The communication patterns we encounter in FV3 are point-to-point, which we realize with MPI.
Each horizontal subdomain only needs to communicate a subset of its results on the boundary to the neighboring subdomains, often called a \textit{halo}.
Halo updates are slightly more complex on the cubed-sphere grid, as data must be transformed according to the orientation of the coordinate system of the adjoining faces of the cube. We thus design a halo updater object in Python that takes care of nonblocking communication, data packing, and transformation based on the pair of ranks.

\subsection{Concessions}

During porting, some concessions had to be made to create an equivalent model that can be statically compiled.

GT4Py does not support absolute indices (e.g., {\small\texttt{a[I - i]}}) and data-dependent horizontal offsets (e.g., {\small\texttt{a[b[0], 0, 0]}}). This led to increase in code size in two instances.
First, in the aforementioned horizontal region computations, some code had to be specialized for each edge/corner. Second, there are modules that behave identically, except for the horizontal direction. For example, identical computations that involve either {\small\texttt{a[1, 0]}} or~ {\small\texttt{a[0, 1]}}. As there is no way to parametrize the dimension as a function argument, these modules had to be duplicated.

Another concession has to do with the GT4Py parallel model. Due to the abstraction of loops, some synchronization points were pre-determined and had to be worked around by splitting stencils into multiple functions.

Even though DaCe is flexible in compiling Python code, we had to limit the behavior of the code between stencils. In particular, we forgo dynamically-sized lists in performance-critical code and Python syntactic features such as introspection (e.g., live state {\small\texttt{\_\_dict\_\_}} updates).
In order to ensure manageable size of the generated program, as a last consideration we explicitly mark loops to be (or not) unrolled.

\section{Hardware Mapping and Orchestration}
\label{sec:mapping}

With the FV3 dynamical core implemented as Python modules that invoke GT4Py stencils, in this work we implement a GT4Py backend that generates SDFGs (the DaCe IR). The goal is twofold: (a) GT4Py stencils can generate fast GPU code via data-centric optimizations; and (b) DaCe can parse the entire dynamical core as one unit, enabling global optimization.

\subsection{From GT4Py to DaCe}

Converting GT4Py stencils to SDFG occurs after GT4Py applies domain-specific optimizations (such as removing unnecessary computations) on the input code.
Upon translation to an SDFG, each stencil computation is represented as a library node. These {\small\texttt{StencilComputation}} nodes
are annotated with attributes to specify their \textit{schedule} for hardware mapping. We define a stencil schedule as follows:
\begin{itemize}
    \item Order of iteration, i.e., along which dimension grid points are accessed with unit stride. 
    \item Tiling and tile sizes in each dimension.
    \item Target for expanded SDFG maps, e.g., GPU thread-block.
    \item Whether iteration in each dimension should be scheduled as a map or a loop. This allows to trade parallelism for better caching of accessed values.
    \item Fields for which data is retained in caches are specified alongside the kind of storage for those caches, such as shared memory or registers.
    \item Horizontal Regions can be implemented as separate maps (i.e., multiple kernels) with an iteration over the respective sub-domain or as a map over the full domain with code predicated on the index for individual statements.
\end{itemize}
The schedule thus defines a general optimization space for a single stencil.
Based on the dependencies between data accesses in each node, only a subset of combinations of these settings is valid. For each node, we are able to generate a list of feasible options from which we make a preferred choice, which can be used for tuning. For some settings, default values are selected such that caches always reside in the fastest memory possible. For example, cached values that are accessed by a single thread will be held in registers, while those that require synchronization between threads are kept in shared memory.

\begin{figure}[t]
\centering
\includegraphics[width=.7\linewidth,page=7]{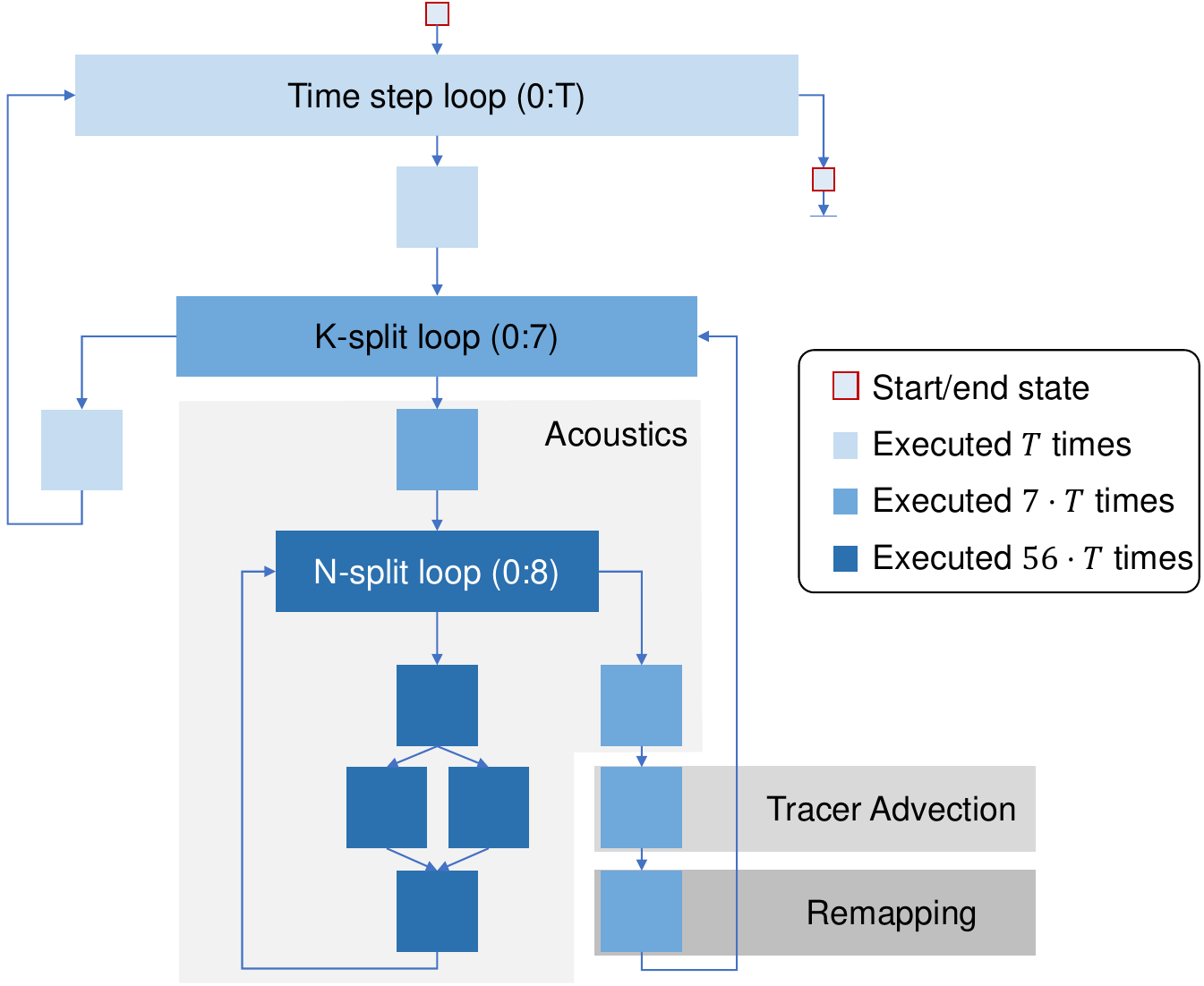}
\caption{Coarse-grain state machine for the configuration of the FV3 dynamical core used in our simulations.}
\vspace{-1em}
\label{fig:cfg}
\end{figure}

When a Python function is parsed by DaCe, calls to GT4Py stencils are detected and insert the resulting library nodes into the graph.
The representation can thus also be used to analyze and optimize stencils on an inter-procedural level, by considering subgraphs with multiple such nodes.

\subsection{Orchestration}

Full-program optimization is a key to striking a balance between performance and productivity. As the Python FV3 implementation defines and orders stencils by their physical meaning (e.g., diffusion or advection) rather than by optimization opportunities, it is beneficial to consider optimizing transformations across multiple stencil invocations. 

Constructing a single SDFG from the entire program creates several performance benefits. It eliminates the Python interpreter's overhead, enables optimizations across functions and translation units (e.g., choosing a data layout to persist across an entire section of the dynamical core, propagating constants into GPU kernels), and allows the DaCe framework to remove unused memory and allocate it outside of the critical path. Additionally, we can guide performance engineering by modeling the performance characteristics (compute and memory bounds) of the entire program.

When orchestrating a code of such magnitude, it is important to ensure productivity is not reduced. By keeping a connection to the Python interpreter via callbacks, our orchestration retains the capability to print out arrays, call external modules during execution (such as plotting libraries), validation, debugging with breakpoints, and use specialized methods for data I/O and boundary conditions.

In the Python FV3 implementation, apart from GT4Py stencils the code uses several design patterns. As mentioned, the code utilizes OOP via classes and fields, uses dictionaries, and includes control flow expressed as loops and branches (some depending on fields and dictionaries). Fig.~\ref{fig:cfg} provides a schematic overview of FV3 as a state machine.

\begin{figure}[t]
\centering
\includegraphics[width=.5\linewidth,page=9]{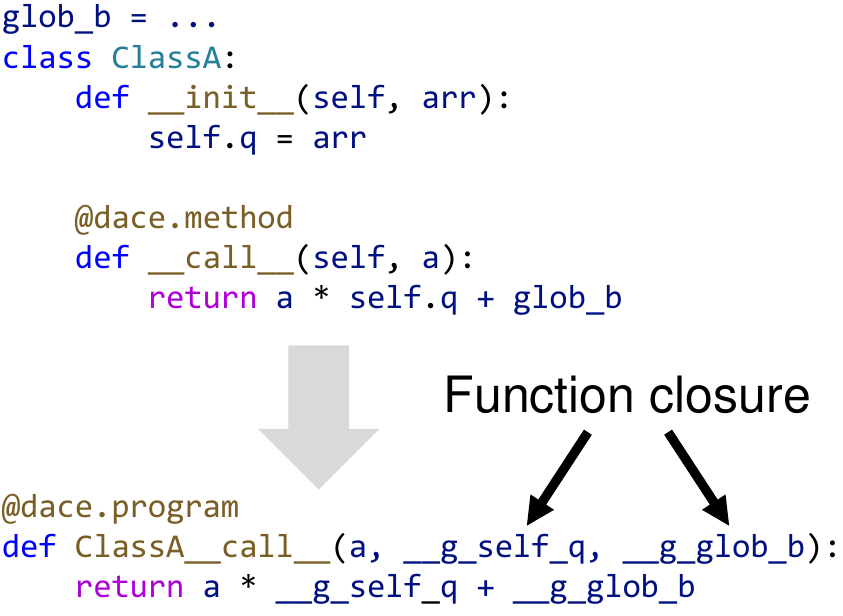}
\caption{Closure resolution for a data-centric method.}
\vspace{-1em}
\label{fig:closure}
\end{figure}

To make the FV3 Python code analyzable w.r.t. data movement, we implement a new Python preprocessor that can handle the aforementioned design patterns:

\begin{itemize}
    \item \textbf{Constant Propagation}: The first step propagates constants forward (i.e., {\small\texttt{i = 5; j = i + 1; b = a[j]}} will become {\small\texttt{b = a[6]}}), performs loop unrolling, and dead code/branch elimination (namely, removing code that will definitely not be executed). This handles cases such as dictionary accesses in a loop (used, e.g., for tracers in FV3). 
    \item \textbf{Closure Resolution}: Methods and functions that depend on external data are converted to free functions (Fig.~\ref{fig:closure}). Resolving closures inlines class structures at preprocessing time, supporting Python OOP. With closures and constants resolved, a call-tree analysis detects and consolidates multiple instances of the same array object (e.g., used in different classes) to avoid data races.
    \item \textbf{Automatic Callbacks}: We extend DaCe to automatically generate callbacks for functions that cannot be parsed as C function pointers. Native Python containers are directly supported, mapping CuPy arrays to GPU pointers and NumPy to CPU. To avoid reordering callbacks during data-centric optimization, we add a \texttt{\_\_pystate} dummy data container as an input and output for each callback, similarly to Calotoiu et al.~\cite{c2dace}.
\end{itemize}

Working with large weather models, scalability becomes a requirement. 
After library node expansion, the SDFG of the orchestrated dynamical core comprises 26,689 dataflow nodes in 3,179 states, grouped into 4,241 unique GPU kernels (maps). Since the model contains loops, some of these kernels are invoked multiple times ($\le$ 56) under different settings. This necessitates a programmatic approach for optimization. 


\section{Data-Centric Optimization}\label{sec:opt}
\label{sec:optimization}

To accelerate the performance of the dynamical core, we realize a disciplined approach for optimization based on model-driven performance engineering~\cite{hoefler-sop-modeling}. Briefly, the application is divided into components on the critical path, and performance bounds (e.g., computational, I/O complexity) are computed on each component and the program as a whole. These serve as guides to how much could each component be optimized further, or hint that a global schedule change should be applied.

\begin{figure}[t]
\centering
\includegraphics[width=\linewidth,page=2]{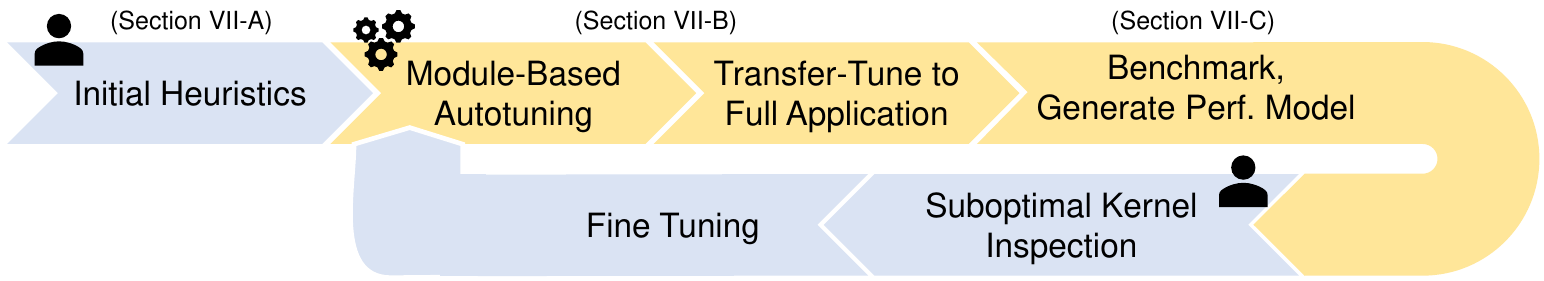}
\caption{Optimization pipeline cycle.}
\vspace{-1em}
\label{fig:methodology}
\end{figure}

Performance engineers are always assumed to be present in a production environment, and this work creates a methodology that reduces the optimization cycle duration by automating as much of the pipeline as possible. As depicted in Fig.~\ref{fig:methodology}, the data-centric approach can be used to mechanize the modeling and tuning parts, which is enabled by the schedule-free declarative approach of GT4Py and the mutable data movement of the SDFG IR. Should the scientists modify the weather model's equations, consider a different model altogether, or upon using a new hardware architecture, the proposed pipeline can be easily reapplied.

The optimization pipeline is composed of four general steps: \textbf{initial heuristics} for scheduling stencils are determined based on estimates and local optimization. Subsequently, \textbf{auto-tuning} is performed on repeating subunits of the application (e.g., individual stencils, modules), and \textbf{transferred} to the full application. Automated performance bounds analyses then guide performance engineers to inspect and manually \textbf{fine tune} bottlenecks in the application, prompting the potential start of the next auto-tuning cycle. In the rest of this section, we provide a detailed account of applying this pipeline to FV3, annotating manual (\mopt) and automatic (\aopt) optimizations.

\subsection{Local optimization and initial heuristics}
\label{sec:optimization:local}

As a first step, the pipeline should optimize each stencil and allocated field locally using the infrastructure available in GT4Py and DaCe. In this work, we formalize the search space for local stencil optimization, and then automatically search it on a representative horizontal stencil and vertical solver separately. The system then applies the resulting scheme \textit{en masse} in the dynamical core, providing a better starting point over the default parameters.

There are four aspects to optimize: scheduling strategy for individual stencils, local storage (specifically for vertical solvers), memory allocation for fields, and computational layout (thread to work mapping).

\subsubsection{\aopt{} Scheduling}
Optimizations that will yield a performance benefit are applied aggressively within GT4Py. Between statements in each stencil, the default fusion strategy combines consecutive intervals in forward and backward solvers into a single map (and GPU kernel), which allows to avoid flushing and re-initialization of cached values to and from global memory between loops. Subsequently, kernel fusion is applied on the thread level if no dependency between threads exists, removing unnecessary inter-thread synchronization. Lastly, consecutive operations in a single \texttt{computation} with dependencies between threads is fused by redundantly computing the results of the first operation (as can be seen in Fig.~\ref{fig:code:sdfg-expanded}).

\subsubsection{\aopt{} Local storage}
The following transformations are applied to avoid load and store operations from or to global memory: (a) temporary fields that are only accessed in a single thread are replaced by local variables; (b) load operations for fields that are overwritten before being read are removed; and (c) values that are used in consecutive iterations of forward and backward solvers need only to be loaded from global memory on their first access, buffered locally on registers during iterations, and flushed back to global memory only when not needed again.

\begin{figure}[t]
\centering
\includegraphics[height=1.6in,page=6]{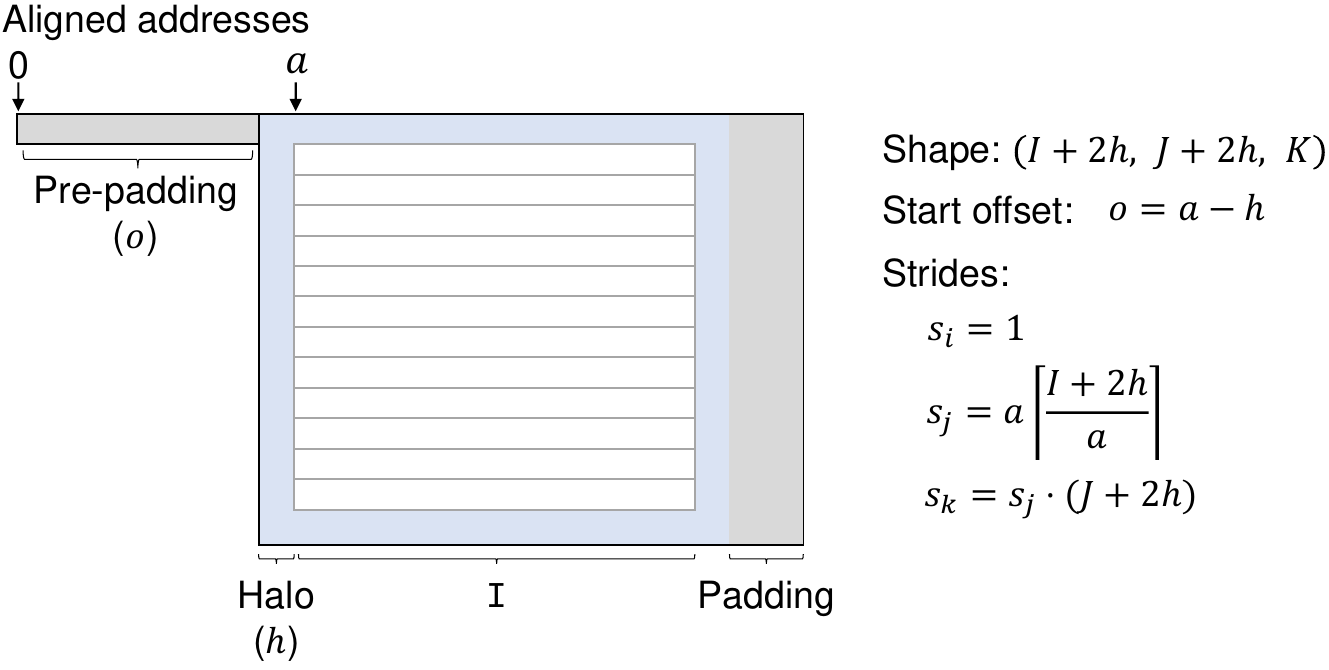}
\caption{Memory allocation scheme for desired alignment $a$.}
\vspace{-1em}
\label{fig:allocator}
\end{figure}

\subsubsection{\mopt{} Memory allocation}
We design a domain-specific optimization space for memory layout of stencil fields.
Allocating the two- and three-dimensional fields used in the model can be parameterized by several knobs. While the shape is fixed, strides can affect layout and padding, whereas pre-padding can be applied to ensure certain elements are aligned.
For our experiments, we use the scheme depicted in Fig.~\ref{fig:allocator}. FORTRAN data layout (\texttt{I}-contiguous) is used since it generates wide loads on the largest dimension (especially vertical solvers, which only access one value of \texttt{K} at a time). To create coalesced writes and reads of fields without offset, we apply pre-padding such that the first non-halo element is aligned, yielding up to 20 $\mu$s ($\sim$5\%) of improvement on the tested stencil.

\subsubsection{\aopt{} Computational layout}\label{sec:opt:complay}
Following an automated sweep of all valid layouts, the results dictate the following schedules. For horizontal stencils: {\small\texttt{[Interval, Operation, K, J, I]}} (last dimension corresponds to {\small\texttt{threadIdx.x}}). For vertical solvers, we use {\small\texttt{[J, I, Interval, Operation, K]}}. These results also correspond to the above data layout and improve coalesced access on the GPU.

\subsection{\aopt{} Transfer Tuning} \label{sec:optimization:transfer}

Auto-tuning provides a convenient optimization framework, as it does not require a hand-crafted performance model.
Unfortunately, exploring the configuration space of transformations for the entire dynamical core is infeasible for the size of FV3.
A typical solution is to only consider a subset of the modules and tune them independently, e.g., the most frequent functions.
However, certain \textit{motifs} recur often in weather and climate codes and can be leveraged to further decrease tuning time.
For instance, if fusing a 5-point stencil and a subsequent element-wise operation turns out to be a positive configuration, it will likely be a positive configuration in other modules, regardless of the actual function.

\begin{figure}[t]
\centering
\includegraphics[height=1.6in]{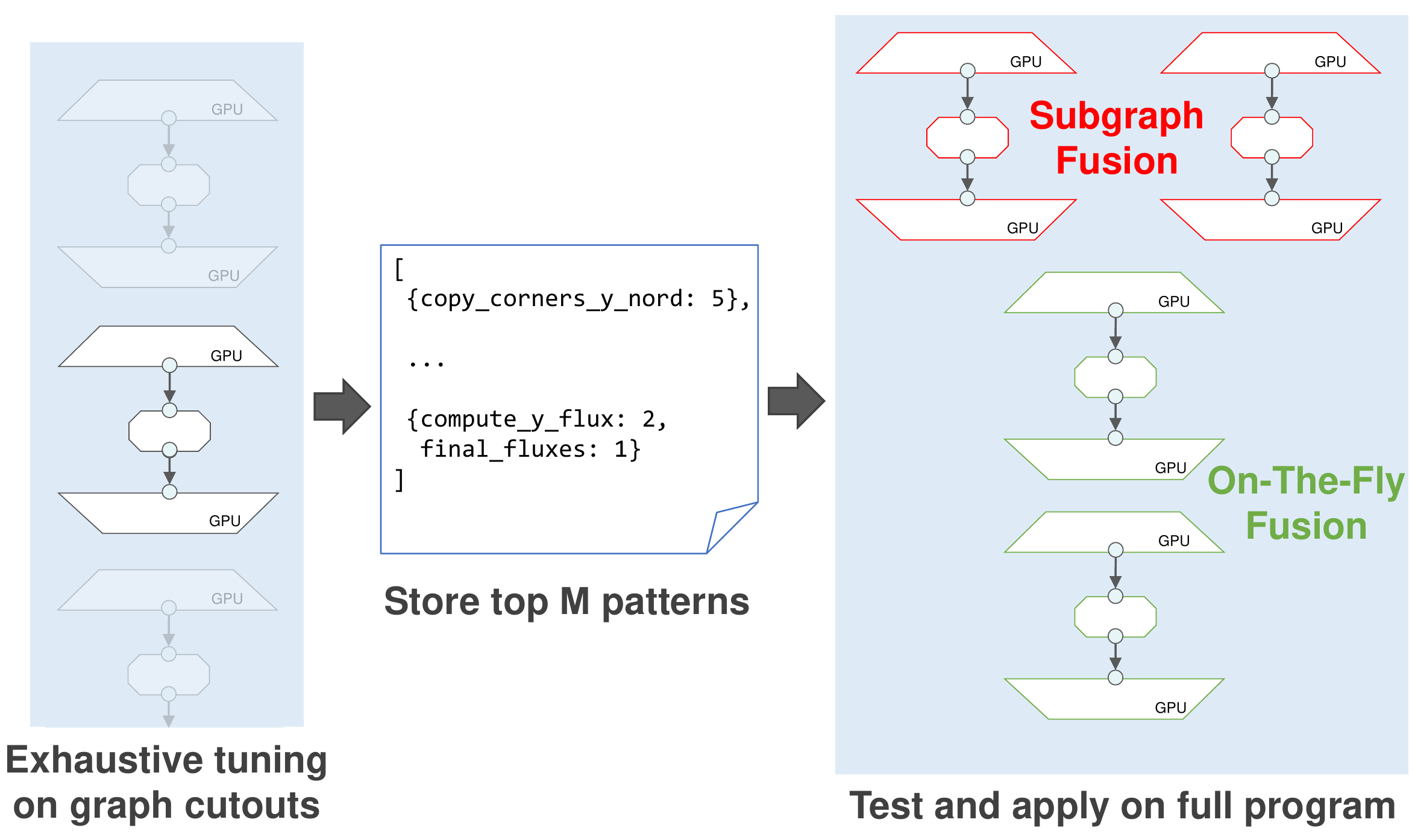}
\caption{Transfer tuning scheme.}
\vspace{-1em}
\label{fig:transfer-tuning}
\end{figure}

We propose a novel method to transfer knowledge across different regions of the code via extracting data movement patterns and reapplying them on other, untuned sections.
Transfer Tuning happens in two phases (Fig.~\ref{fig:transfer-tuning}):
in a first phase, the SDFG of the full program is divided into a set of ``cutout'' subgraphs, each of which is tuned individually.
The best $M$ configurations are translated into optimization patterns and tested on the whole graph in the second phase.
The second phase ensures that found patterns are only applied if they also provide a local performance improvement on a match.

The second phase of transfer tuning requires a description of the patterns that can be searched for in the rest of the graph.
We define a configuration by a set of names of the participating stencils and which transformations were applied, but other implementation-agnostic graph motifs could be used.

\subsubsection*{Case Study: Finite Volume Transport}
We tune the \texttt{fv\_tp\_2d} module, which recurs often in FV3 (Fig.~\ref{fig:fv3}). There are $127$ possible graph ``cutouts'' in the module, and optimization patterns found in it generalize to other parts of FV3.
The maximum number of configurations for a cutout in this module is $48$ and the total number of configurations is $1$,$272$, which are searched exhaustively in the first phase.

For the transfer, the best ($M=2$) configurations of each cutout are considered on the target graph, which is the full dynamical core.
In order to prune the space of matches, we only consider the first match for each pattern in each state, and only match the most performance-improving pattern. In total, 603 transformations were automatically found and transferred to the target FV3 graph.
Without transfer tuning, the number of configurations to consider would be $\ge 30$,$302$,$185$.

On a single node of the Piz Daint supercomputer, the runtime of the first phase of transfer tuning on the module takes 2:42 hours for pattern extraction.
The second phase applied on FV3 takes 8:24 hours.
Thus, auto-tuning the entire dynamical core can run in feasible time and drastically reduce the global optimization search space.

\subsection{\aopt{} Memory-bound analysis and \mopt{} fine-tuning}

\begin{figure}[t]
\centering
\includegraphics[width=\linewidth]{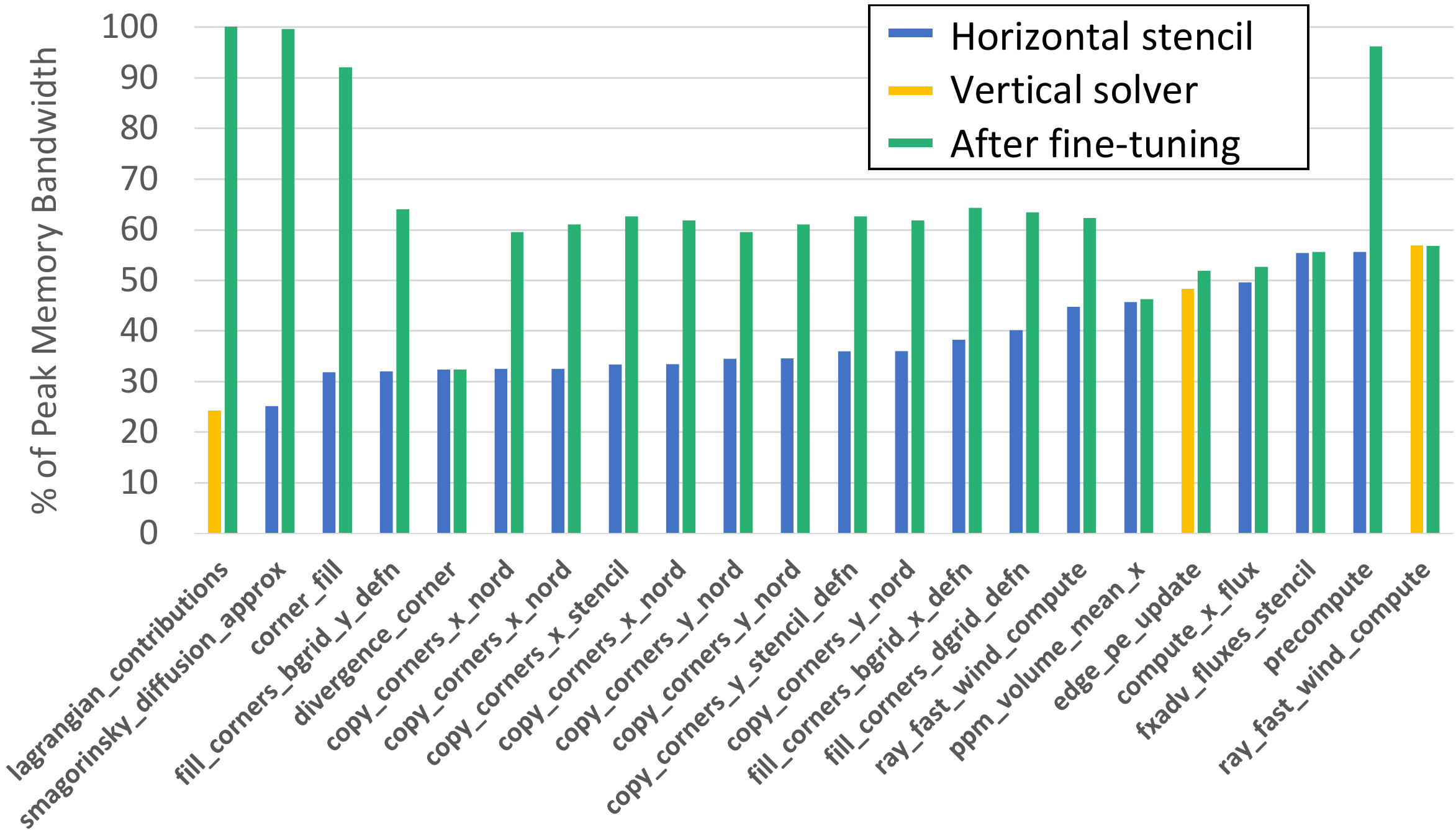}
\caption{Model-augmented kernel runtimes.}
\vspace{-1em}
\label{fig:bounds}
\end{figure}

One of the advantages of using a data-centric IR is the capability to perform \textit{automated} performance modeling. We use modeling, combined with runtime results, to create an overview for the performance engineer, breaking down which parts are most important to optimize further.

We write a simple, general script (17 lines of Python code) that computes the peak performance of each stencil, if it were memory bandwidth bound. The script avoids modeling L1 and L2 caches by considering every element of the field being accessed once, even if multiple threads access the same element.
Since kernels execute under different configurations (due to array sizes or differently masked elements), performance engineers see the maximal reported runtime and largest modeled configuration for performance bounds analysis. To rank kernels by overall importance, the results are sorted by summarized runtimes grouped by kernel type. 

In Fig.~\ref{fig:bounds}, we list the results of the worst-performing, most important kernels from our initial (i.e., first cycle) performance model. In subsequent cycles, as a result of further tuning, most of the shown kernels are above 60\% peak memory bandwidth. 
Below, we showcase some of the manual fine-tuning efforts.
The rest of the optimizations are outlined in Section~\ref{sec:results:opt}.

\subsubsection{\mopt{} Case Study: Region Pruning}
Fig.~\ref{fig:bounds} shows that many of the kernels named \texttt{"*corner*"} under-perform based on the model. Upon performance engineer inspection, it was found that horizontal regions (Section~\ref{sec:regions}) are the root cause. Specifically, computations such as the following:

{
\footnotesize
\begin{lstlisting}
with horizontal(region[i_start - 1, j_start - 1]):
    q = sw_mult * q_corner[0, 1, 0]
\end{lstlisting}
}

\noindent would generate code that spawns many more threads than moved data elements, especially on ranks that do not require computation in the corners. As a result, many of the spawned GPU threads remain idle.

To resolve this, the performance engineers wrote a specialized DaCe transformation that detects such discrepancies via pattern-matching in GT4Py and (a) eliminates unused statements in regions based on MPI rank; and (b) generates smaller GPU kernels that do not spawn idle threads. With that, many of the corner/edge kernels run twice faster, but still not able to fully utilize the GPU memory bandwidth, likely due to the small data size processed.

\subsubsection{\mopt{} Case Study: Smagorinsky Diffusion}
As the second-slowest kernel in Fig.~\ref{fig:bounds}, Smagorinsky Diffusion demonstrates the ease of finding performance discrepancies with our model. The kernel appears to under-perform albeit not containing any specialized I/O complexity patterns nor excessive floating point operations. Upon manual inspection of its source code, the stencil uses the power operator extensively:

{
\footnotesize
\begin{lstlisting}
with computation(PARALLEL), interval(...):
  vort = dt * (delpc ** 2.0 + vort ** 2.0) ** 0.5
\end{lstlisting}
}

The resulting generated code contained {\small\texttt{pow(delpc, 2.0)}} and {\small\texttt{pow(..., 0.5)}}, which are general-purpose and thus highly inefficient. We thus created a transformation that converts powers of positive and negative integers, as well as 0.5, into multiplication loops and {\small\texttt{sqrt}} respectively. Following the transformation, the kernel runtime is reduced from 511.16 $\mu$s to 129.02 $\mu$s, with the automated model reporting 99.68\% utilization. The runtime of other kernels improved as well, resulting in a 1.81\% overall speedup for each time step.

\begin{table*}[t]
\centering
\caption{Performance Analysis of Representative FV3 Modules}
\footnotesize
\setlength{\tabcolsep}{3pt}
\begin{tabular}{lrrrrrp{1em}rrrrr}
\toprule
& \multicolumn{5}{c}{Riemann Solver} && \multicolumn{5}{c}{Finite Volume Transport}\\
\cmidrule(l){2-6}\cmidrule(l){8-12}

            & \multicolumn{2}{c}{FORTRAN} & \multicolumn{2}{c}{GT4Py+DaCe} &      &       &\multicolumn{2}{c}{FORTRAN} & \multicolumn{2}{c}{GT4Py+DaCe} \\
            \cmidrule(l){2-3}\cmidrule(l){4-5}\cmidrule(l){8-9}\cmidrule(l){10-11}
Domain Size (relative size) & Time [ms]    & Scaling  & Time [ms]       & Scaling & Speedup     & & Time [ms]    & Scaling  & Time [ms]       & Scaling & Speedup       \\  
\midrule
128$\times$128$\times$80 (1x)             & 12.27    & ---     & 1.85            & --- & 6.63$\times$       &     &3.41          & ---         & 1.81             & ---           & 1.88$\times$    \\
192$\times$192$\times$80  (2.25x)          & 27.94    & 2.28     & 3.86            & 2.08 & 7.25$\times$     & & 12.31         & 3.61        & 3.41             & 1.88          & 3.61$\times$    \\
256$\times$256$\times$80  (4x)             & 52.40    & 4.27     & 6.96             & 3.76 & 7.53$\times$    & & 35.79         & 10.49       & 5.67             & 3.13          & 6.31$\times$    \\
384$\times$384$\times$80  (9x)             & 121.80   & 9.92     & 15.31            & 8.26 & 7.96$\times$   & &106.66        & 31.27       & 13.10            & 7.23          & 8.14$\times$    \\
\bottomrule
\end{tabular}
\vspace{-1em}
\label{fig:pbounds}
\end{table*}

\section{Experimental Setup}
\label{sec:exp}

In the following sections, we measure the performance of the Python FV3 implementation and estimate the efforts to apply the methodology to other models or hardware.

All experiments were conducted on the Swiss National Supercomputing Centre's Piz Daint supercomputer. The cluster contains 5,704 Cray XC50 nodes, each with an Intel Xeon E5-2690 v3 12-core CPU, NVIDIA Tesla P100 GPU (16 GB RAM), and 64 GB of host RAM. The nodes are connected via the Cray Aries interconnect.
Each experiment is performed at least 10 times and the median result is reported.

We use Python 3.8.2, DaCe version 0.13, GT4Py revision \texttt{6588aa8}, fv3gfs-fortran revision \texttt{9030cf9}, and PAPI 6.0.0.9. Python FV3 was compiled with CUDA 11.2 and GCC 9.3.0, whereas the FORTRAN FV3 was compiled with Intel IFORT version 19.1.3.304. For all experiments, unless otherwise stated, we use a domain size of $192\times 192$ horizontal grid-points per compute node and $80$ vertical levels with double-precision elements. Using this setup on all of the nodes of Piz Daint would result in a global simulation of 1.5$\,$km resolution. Our performance baseline is the optimized FORTRAN version of FV3, which is tuned for multicore CPUs. The FORTRAN version's configuration was tuned for production runs, at 6 ranks per node with 4 threads in each rank, thereby utilizing hyperthreading on the 12 physical cores.

\section{Performance Bounds} \label{sec:bounds}

Following the model-driven performance engineering discipline, we begin by characterizing the workload.
On hardware architectures where memory movement is orders of magnitude more expensive than computation, stencil programs will be memory-bound. This stems from the fact that operations in a stencil are typically proportional to the number of accesses. This is also the case in FV3. We use PAPI~\cite{papi} to measure the FORTRAN dynamical core and observe that 40.15\% of the executed instructions were load/store operations.


\subsection{Memory bandwidth}
For our target local domain size, we claim that FV3 is in particular a memory bandwidth-bound workload.
To empirically prove this claim, as well as understand the expected speedups of GPU vs. CPU, we measure the peak and maximum attainable memory bandwidth of the two architectures. 

We use the STREAM benchmark for the CPU~\cite{stream} and the CUDA memory bandwidth test~\cite{cuda-samples} for the GPU. The reported numbers for Piz Daint are 43.77 GB/s for the Haswell CPU, and 501.1 GB/s for the Pascal GPU. To verify that GT4Py and DaCe can achieve maximal bandwidth, we run a ``copy stencil'' (one input, one output) on the target domain size. We measure a CPU bandwidth of 40.99 GiB/s and a GPU memory bandwidth of 489.83 GiB/s. This indicates that the sizes are large enough to sustain the full bandwidth, as well as expect a maximum speedup of 11.45$\times$ for a memory-bound problem.

\subsection{Vertical solvers: Riemann Solver}

In the acoustic substep, the \texttt{riem\_solver\_c} module solves for the nonhydrostatic terms of vertical velocity and pressure perturbation~\cite{chen_control-volume_2013, chen_towards_2018}. We use this module as a representative of vertical solvers, which typically do not perform well in the FORTRAN FV3 \texttt{K}-blocking schedule. 

The semi-implicit method used for discretization is implemented using a tridiagonal solver, which is divided into three GT4Py stencils and scheduled as 22 GPU kernels. Table~\ref{fig:pbounds} (left) lists performance measured on several domain sizes, as well as the relative scaling w.r.t. the ratio of grid points. 

The table exhibits two scaling trends. On the one hand, the FORTRAN version scales increasingly worse as the domain size grows. Scaling worse than the ideal scaling indicates that the CPU caches no longer suffice for larger domains. Moreover, the increase in gap between the slowdown and domain size suggests that the schedule is suboptimal. Different data layouts (e.g., \texttt{K}-contiguous) could mitigate this effect, but without a DSL approach that would constitute an intrusive code modification for every stencil that accesses those fields.

The second scaling trend can be seen in the data-centric Python version. 
We see that the slowdown is always smaller than the domain size scaling. This typically indicates that not enough parallelism is exposed on the smaller domain sizes (only 2D thread grids are used). This can also be confirmed by the gap decreasing as the domain size increases.
Overall, for all sizes starting from the target domain size, the speedup is over 7.25$\times$ over the FORTRAN version. The speedup remains relatively stable, and is expected to increase further (albeit slightly) with larger domain sizes.

\subsection{Horizontal stencils: Finite Volume Transport}

An integral part of FV3 is the \texttt{fv\_tp\_2d} module, which is a subroutine to compute fluxes for horizontal finite volume transport \cite{putman_finite-volume_2007, lin_multidimensional_1996}. 

In FORTRAN, the module is designed to be two-dimensional and mass-conserving, i.e., each call operates on a horizontal slice of the transported scalar, and there are no dependencies between the vertical levels. To achieve high performance efficiency, vertical \texttt{K}-blocking is employed. 
This schedule and blocking strategy make this module challenging to gain speedups over, as we observe that CPU caches are highly utilized ($\sim$0.13\% of the load/store instructions end up as L3 cache misses on our target domain size).

The runtimes in Table~\ref{fig:pbounds} (right) indicate that the FORTRAN version relies on CPU caches. With each domain size increase, the slowdown scales at a higher factor. However, the gap between slowdown and ideal scaling is decreasing, further indicating that we are outside of the cache capacity regime. We can also see that the GPU is again underutilized, with runtime scaling at a lower factor than the number of grid points. As the CPU starts to access off-chip memory, we see that the speedups increase towards the bandwidth ratio.

Another way to look at this performance behavior is as a hardware/software co-design guideline. An ideal hardware architecture for sequences of horizontal stencils would have the off-chip memory bandwidth and core count of a GPU, but could retain cached memory \textit{across kernels}, benefiting from the inter-stencil memory reuse similarly to a CPU.

\section{Distributed Performance}
\label{sec:distributed}

For our test case we set the initial state of the model corresponding to a uniform zonal flow with a perturbation which evolves into a baroclinic instability~\cite{ullrich2014proposed}. This analytical test case enables generation of arbitrary domain sizes and a fast visual verification of the results.

\subsection{Optimization pipeline}\label{sec:results:opt}

\begin{table}[t]
    \centering
    \setlength{\tabcolsep}{4pt}
    \caption{Dynamical Core Optimization} \footnotesize
\begin{tabular}{llrrr}
\toprule
Cycle & Version & Date & Time [s] & Speedup \\
\midrule
               & FORTRAN                     & ---     & 16.36 & 1.00$\times$             \\
               & GT4Py + DaCe (Default)      & Feb 15  & 10.87 & 1.50$\times$             \\
               \midrule
Cycle 1        & Stencil schedule heuristics & Mar 13 & 5.56  & 2.94$\times$             \\ 
               & Local caching               & Mar 15  & 5.45  & 3.00$\times$             \\ 
               & Optimize power operator     & Mar 17 & 5.35 & 3.06$\times$             \\ 
               & Split regions to multiple kernels & Mar 18  & 4.82 & 3.39$\times$             \\\midrule 
Cycle 2        & Lagrangian contrib. reschedule    & Mar 25  & 4.82 & 3.40$\times$             \\ 
               & Region pruning              & Mar 31  & 4.77  & 3.43$\times$            \\ 
               & Transfer Tuning       & Mar 31  & 4.61  & \textbf{3.55$\times$}    \\ 
\bottomrule
\end{tabular}
    \label{tab:opts}
    \vspace{-1em}
\end{table}

The smallest distributed run configuration uses 6 nodes, where each node takes a single face of the cubed sphere. In this configuration, each node computes all of the special computations required at tile edges and corners. We use this 6-node run as a case study for productivity and performance improvement over the course of optimization. The results and dates of each version are summarized in Table~\ref{tab:opts}.

The optimization consisted of two performance engineering cycles over a six week period (including tooling development) by four developers, where we follow the optimization cycle laid out in Fig.~\ref{fig:methodology}. 
Once the modeling/tuning tools were developed, and the search spaces were defined, most optimizations took days to perform. The cycle informed the developers on which parts of the models to focus, and the parametrization enabled modifying individual components with ease.

More specifically, as the default schedules were suboptimal, auto-tuning stencils locally to obtain heuristics for scheduling (Section~\ref{sec:opt:complay}) yields a 1.96$\times$ speedup. As the cycle progresses with model-driven fine tuning, optimized kernels yield smaller but non-negligible returns. At the end of the second cycle, we run another tuning pass on the same day, transferring the results from \texttt{fv\_tp\_2d} to the rest of the dynamical core for a 3.47\% speedup. Additional cycles could improve the performance further (albeit with diminished returns), or reapplied to other hardware as discussed in Section~\ref{sec:porting}.

It is important to keep in mind that \textit{all} performance engineering was accomplished without modifying the user-code, but by applying optimizations in the toolchain.

\subsection{Performance at scale}\label{sec:results:scale}

\begin{figure}[t]
\centering
\includegraphics[width=.9\linewidth]{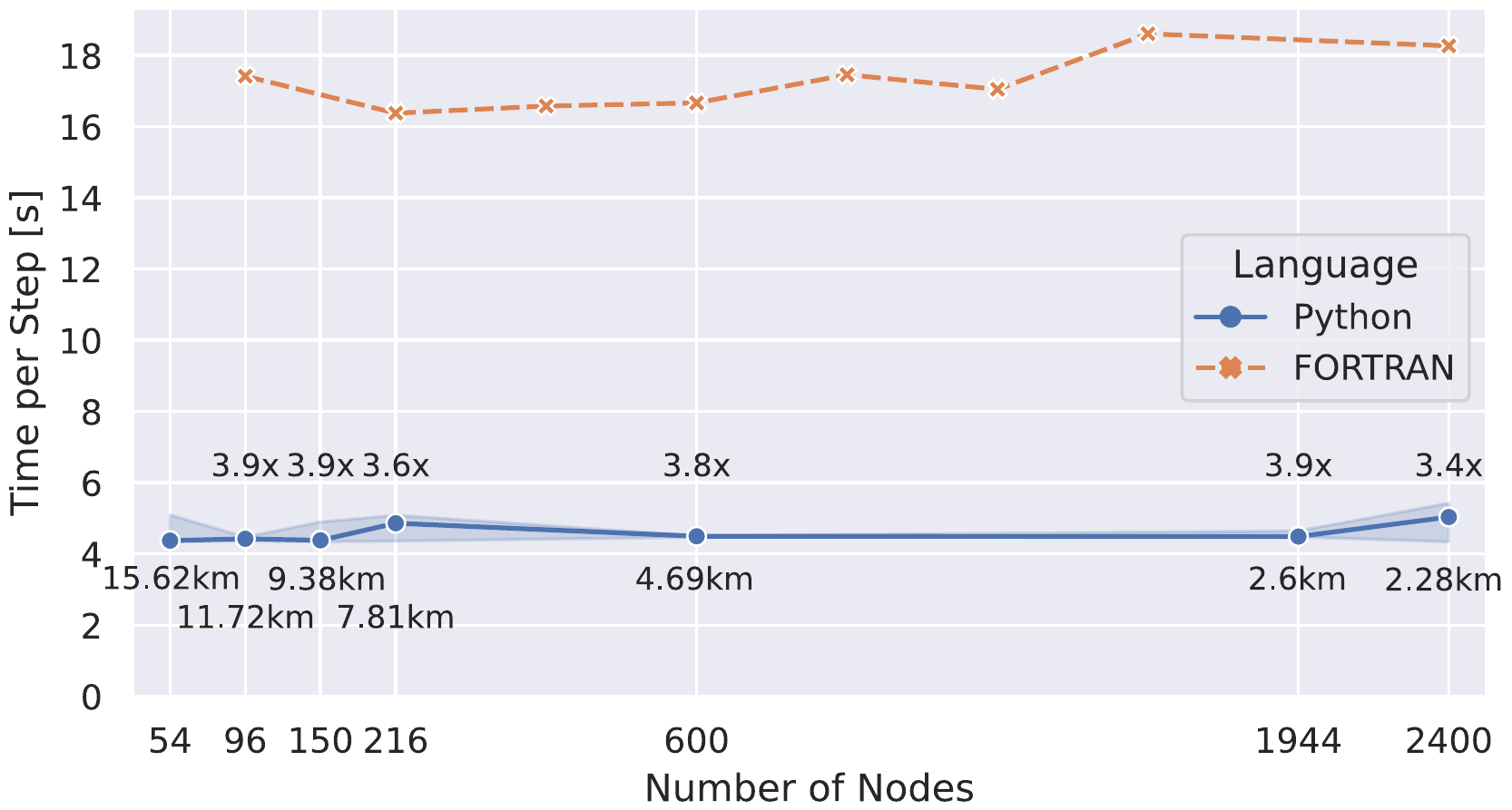}
\caption{Large-scale performance of FV3.}
\vspace{-1em}
\label{fig:scaling}
\end{figure}

We run global simulation experiments, scaling from 54 nodes (15.6$\,$km grid spacing) to 2,400 nodes (2.28$\,$km grid spacing) and keeping the same grid size per process. The time per timestep is plotted in Fig.~\ref{fig:scaling}, showing the weak scaling of the Python FV3 vs. the FORTRAN version, where shaded regions denote the nonparametric 95\% confidence intervals.

As expected, the speedups at scale are marginally higher (up to \textbf{3.92$\times$}) than the ones obtained on the 6-node run.
We observe a throughput of 0.11~simulated years per day (SYPD) for the 2.28$\,$km simulation, and nearly perfect weak scaling (as per-node communication remains similar). This is all achieved with \textit{object-oriented high-level declarative Python code}, which can be easily modified and further developed.

To demonstrate the portability of our approach, we also run the code unmodified on the JUWELS Booster supercomputer, which consists of multi-GPU nodes with NVIDIA Tesla A100 GPUs. With 54 ranks (14 nodes), we attain a performance of 1.93 seconds per timestep. It is 2.42$\times$ faster than the Piz Daint result, which is promising given that the memory bandwidth of A100 is 2.83$\times$ faster than the P100 GPU~\cite{p100datasheet,a100datasheet}.


\section{Productivity}
\label{sec:porting}

The effort to port the dynamical core to Python took approximately two person-years.
This includes not only the numerical operators, but also the initialization code, driver routines, and testing infrastructure.
With the tooling now implemented, future models should be easier to port.
Once the first version of the Python model validated against the reference FORTRAN model, the optimization efforts commenced.

Our approach for performance engineering is, by design, portable to different hardware architectures.
However, re-optimization time is dependent on the target architecture type.

Table~\ref{tab:workflow} summarizes which elements of our methodology need to be repeated.
\textit{Compile} refers to the need to generate and recompile the code for a specific hardware target and is done with a single re-run of the model.
\textit{Tuning} refers to the transfer tuning approach described in Section~\ref{sec:optimization:transfer}.
\textit{Optimization cycles} refer to estimating the times the modeling-inspection-tuning cycle from Fig.~\ref{fig:methodology} should be repeated, in order to achieve similar \%peak performance. 
Lastly, \textit{code generator} refers to writing a backend for DaCe, if there was no prior one.

While the P100 GPU was the target of this effort, performance on the A100 GPUs (Section~\ref{sec:results:scale}) was generated by just recompiling the code (i.e., without further tuning or an optimization cycle). When a different GPU architecture (e.g., AMD) is used, the heuristics and transformations would need to be re-tuned for hardware specifics such as warp size. 

For CPUs, we know from FV3 that some of the GPU schedules are suboptimal for CPU, due to cache behavior and different instruction sets (e.g., ARM SVE). We thus estimate that another optimization cycle or two would be additionally necessary to ensure the best performance.

\begin{table}[t]
    \centering
    \setlength{\tabcolsep}{4pt}
    \caption{Workflow Porting Effort Estimation}\vspace{-0.5em} \footnotesize
\begin{tabular}{lcccc}
    \toprule
    Porting & Compile & (Transfer) & Optimization  & Code \\
            &         & Tuning    & Cycles  & Generator\\
    \midrule
    NVIDIA GPU             & \checkmark  &   & 0 &              \\
    Other GPU              & \checkmark          & \checkmark & 0 &             \\
    CPU  & \checkmark         & \checkmark         & $\approx$ 1--2 &             \\
    Other chip     &    \checkmark      & \checkmark         &  $\ge$ 2        & \checkmark             \\
    \midrule
    New model, same GPU  & \checkmark & \checkmark & $\approx$ 1 &\\
    \bottomrule
\end{tabular}
    \label{tab:workflow}
    \vspace{-1.5em}
\end{table}

\section{Conclusion}
\label{sec:conclusion}

We present a disciplined approach to optimizing weather and climate applications. Using FV3 as a case study, we demonstrate how taking a declarative approach to defining the underlying equations enables full freedom to optimize each stencil locally, and reschedule the model as a whole. Powering this flexible optimization is a data-centric representation, which allows performance engineers to automatically establish performance bounds and optimize computation and data layout, without modifying the source code.

The shown pipeline results in a successful porting of FV3, which is specifically designed and tuned for multi-core CPUs, to run on GPUs at scale, all while taking less than a half of the original code. 
The methodology presented here is applicable to any weather model and various hardware architectures, where it could yield similar, if not higher performance improvements.

\section*{Acknowledgment}
This project received funding from the European Research Council \includegraphics[height=1em]{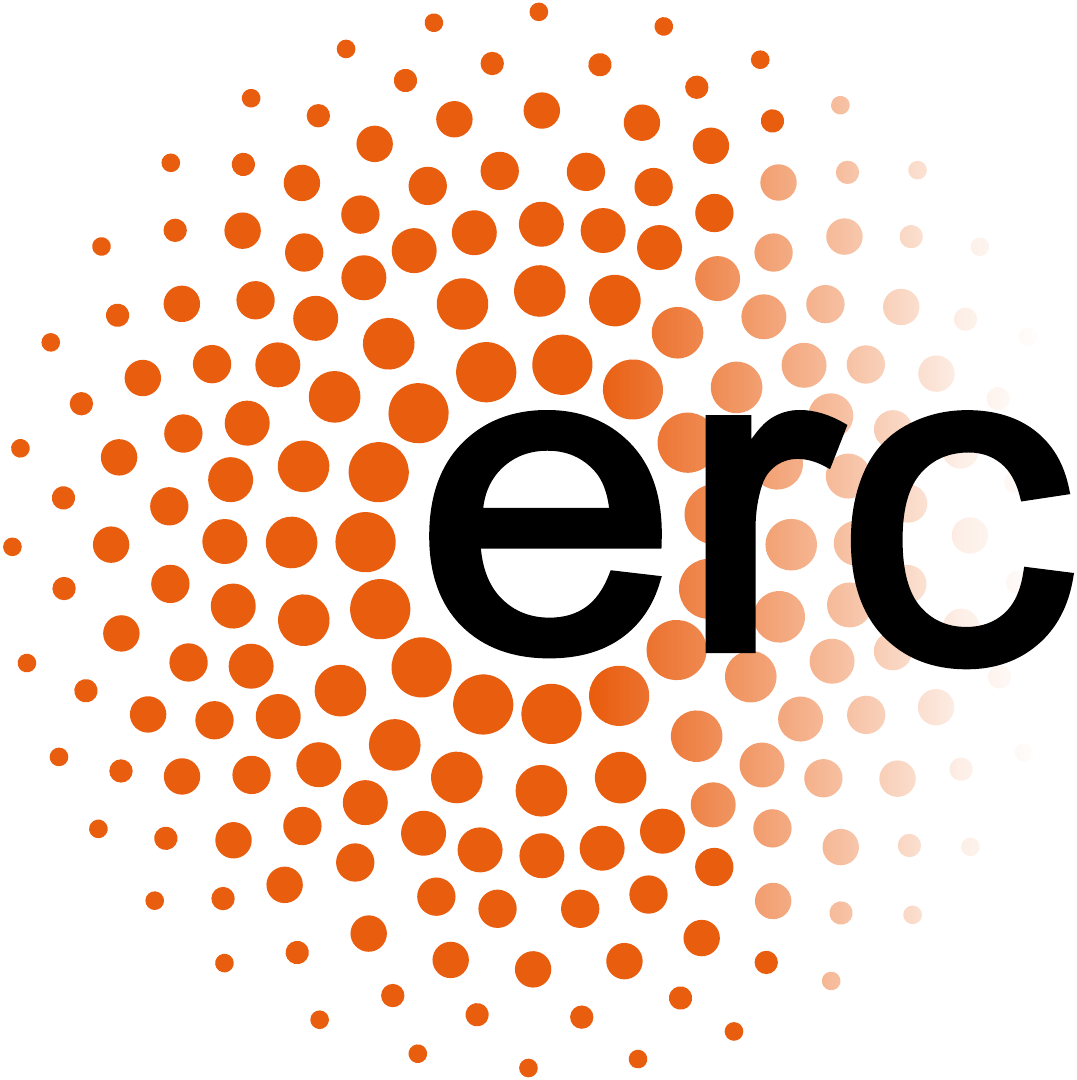} under the European Union’s Horizon 2020 programme (Project PSAP, No. 101002047); and EuroHPC-JU funding under grants DEEP-SEA, No. 955606 and MAELSTROM, No. 955513, with support from the Horizon 2020 programme.
The authors also wish to acknowledge the support from the PASC program (Platform for Advanced Scientific Computing) for the DaceMI project.
T.B.N. is supported by the Swiss National Science Foundation (Ambizione Project \#185778).
We acknowledge contributions from the whole GT4Py team, specifically Hannes Vogt (CSCS) and Enrique Gonzalez (CSCS), for their help in implementing a validating version of FV3 using GT4Py. We also acknowledge the contributions of Christopher Kung (NASA) for his contributions to the port of FV3 to GT4Py. Furthermore, we acknowledge the helpful discussions with Lucas Harris (GFDL) and Rusty Benson (GFDL) to understand the inner workings of FV3. This work was supported by a grant from the Swiss National Supercomputing Centre (CSCS) under project ID s1053. We would like to thank the Allen Institute for Artificial Intelligence (AI2) and Vulcan Inc. for their financial support for part of the staff of this project.

\bibliographystyle{IEEEtran}
\bibliography{references}

\end{document}